\documentclass[prl,twocolumn]{revtex4}
\usepackage{epsfig}
\usepackage{graphicx}
\usepackage{amsmath}
\usepackage{amsfonts}
\usepackage{amssymb}%
\begin{document}
\title{Unraveling the Jahn-Teller effect\\
in Mn doped GaN}
\author{A. Stroppa}
\affiliation{Faculty of Physics, University of Vienna,
and Center for Computational Materials Science,
Universit\"at Wien, Sensengasse 8/12, A-1090 Wien, Austria}
\author{G. Kresse}
\affiliation{Faculty of Physics, University of Vienna,
and Center for Computational Materials Science,
Universit\"at Wien, Sensengasse 8/12, A-1090 Wien, Austria}

\begin{abstract}
We present an ab-initio study of the Mn substitution for Ga in GaN
using the Heyd-Scuseria-Ernzerhof hybrid functional (HSE). Contrary
to semi-local functionals, the majority Mn t$_{2}$ manifold splits
into an occupied doublet and an unoccupied singlet well above the
Fermi-level resulting in an insulating groundstate, which is further
stabilized by a sizeable Jahn-Teller distortion. The predictions are
confirmed using  $GW$ calculations and are in agreement with
experiment. A transition from a localized to a delocalized 
Mn hole state is predicted from GaN to GaAs.
\end{abstract}

\maketitle

\label{introduction}\label{intro}

Semiconductor based spintronics aims to develop  hybrid devices that
could perform all three  operations, logic, communications and
storage within the same materials
technology\cite{intro:spintronics1}. Dilute Magnetic Semiconductors
(DMSs) represent the most promising materials, and undoubtedly,
transition metal doped III-V semiconductors are presently the
workhorse for
spintronics\cite{intro:spintronics2}.
 
Ab--initio simulations based on density functional theory
 have played an important role
in  investigating the physics of
DMSs\cite{intro:DMS_DFT2,intro:DMS_DFT3}.
Nevertheless, the theoretical understanding has been hindered by the
well-known deficiencies of the spin-polarized  local density
approximation (SLDA) and generalized gradient approximation (SGGA)
to the exchange-correlation functional\cite{intro:failures}: the
non-locality of the screened exchange interaction is not  taken into
account and the electrostatic self-interaction is not entirely
compensated. This lack of compensation causes fairly large errors
for localized states, \emph{e.g.} the Mn $d$ states. It destabilizes
the orbitals and decreases their binding energy, leading to an
over-delocalization of the charge density\cite{NatMat}. Another
closely related issue is that the Kohn-Sham gap is usually
a factor 2-3 smaller than the fundamental gap of the
solid\cite{intro:failures}. Whenever the energy position of the
defect level with respect to the Valence Band Maximum (VBM) is
comparable with the Kohn-Sham gap, \emph{e.g.} deep acceptor levels
introduced by substitutional Mn in GaN (Mn$_{\rm
Ga}$)\cite{intro:Mn_Ga}, the calculation of the thermodynamic
transition levels becomes difficult, since all predicted
thermodynamic transition levels are strictly bound by the Kohn-Sham
one electron gap. Although the underestimation of the one electron
gap would even occur for the exact  Kohn-Sham functional, 
 discontinuities in the potential upon adding or removing electrons
correct for this error\cite{intro:failures,ScuseriaReview}. For approximate
functionals, which lack any such discontinuities--- this includes all
available semi-local and hybrid functionals
---agreement between the Kohn-Sham gap and experimental fundamental
gap is a prerequisite for modelling thermodynamic transition
levels and band structure related properties\cite{intro:failures}.

Hybrid Hartree-Fock density functionals\cite{intro:hybrid1} overcome
the two limitations discussed above to a large
extend\cite{intro:failures}. Here, we apply the
Heyd-Scuseria-Ernzerhof (HSE) hybrid
functional\cite{computational:hse} to study the Mn impurity in a GaN
semiconductor host.  Extensive studies of the performance of the HSE
functional in solid state systems can be found in Refs.\
\onlinecite{hsestudies1,hsestudies2,computational:epsilon},
unequivocally showing that hybrid functionals outperform semi-local
functionals for materials with band gaps. The HSE results are
confirmed by $GW_{0}$ calculations, which are the
benchmark method for the prediction of quasiparticle (QP)
energies\cite{computational:GW1}. We will show that the electronic
properties  are accurately described by both methods. The $t_{2}$
manifold is split into an occupied doublet and an unoccupied singlet
giving rise to a \emph{symmetry-broken insulating} ground state
that naturally couples with the ionic lattice, distorting the
otherwise ideal tetrahedral environment of the Mn ion (Jahn-Teller
effect). The calculated thermodynamic transition level
$\epsilon(0/-)$ is in good agreement with experiments. Remarkably,
most  of these features are not captured by standard SLDA or SGGA,
without introducing ad-hoc corrections, such as 
self-interaction corrections\cite{Sic-schemes1,Sic-schemes2}
 or LDA+U corrections\cite{LDA_U}. Hybrid functionals have a single
parameter (non-local exchange) that is once and forever fixed to 1/4
on theoretical grounds\cite{perdew}. 
Furthermore, Stengel \textit{et al.} have very recently highlighted 
some drawbacks of SIC-schemes, concluding that hybrid functionals 
represent the most promising route to reduce self-interaction problems 
while preserving unitary invariance\cite{Nicola}.

The calculations were performed using the projector augmented-wave
(PAW) method\cite{computational:paw} with the Perdew-Burke-Ernzerhof
(PBE) GGA functional\cite{computational:pbe} and
Heyd-Scuseria-Ernzerhof (HSE) hybrid
functional\cite{computational:hse} recently implemented in the VASP
code\cite{computational:vasp}, following exactly the prescription
given in Ref. \onlinecite{krakau} (HSE06). The Ga $3d$ and Mn $3p$
electrons were considered as valence electrons. A soft nitrogen PAW
potential was used and  the energy cutoff was set to 280 eV.
Supercells with  64 and 128 atoms were used with lattice constants
fixed to the optimized HSE value for the bulk crystal.
Brillouin-zone integration was carried out using 8$\times$8$\times$8
and 2$\times$2$\times$2 Monkhorst-Pack grids for bulk GaN and the
supercells, respectively. We used the Van der Walle and Neugebauer
approach\cite{computational:VandeWalle} for calculating the
transition level of Mn$_{\rm Ga}$  using a 128 atom cell. For each
charge state, the atomic positions were relaxed. Errors due to the
electrostatic interactions were taken into account through the
Madelung energy of point charges $q$ in an effective medium with a
static dielectric constant $\varepsilon_{\infty}=5.1$,
$\Delta$E$_{1}$=$q^{2}\alpha/2\varepsilon_{\infty}L$, where $L$ is
the distance between the Mn and its periodic replica, and $\alpha$
is the Madelung
constant\cite{computational:errordefects1}.
$\varepsilon_{\infty}$ was evaluated according to
Ref.~\onlinecite{computational:epsilon}. For the $GW_0$
calculations, we determined the screening properties entering $W_0$
using the random phase approximation (RPA) and PBE wavefunctions and
eigenvalues, but iterated the eigenvalues in $G$ until
selfconsistency was reached. The initial wavefunctions and
eigenvalues in $G$ were determined using the HSE functional. For
bulk materials, this procedure yields results that are within  5\%
of experiment and the selfconsistent sc$GW$ procedure with vertex
corrections that was used in Ref. \cite{bulkGaN:Kr3}. The latter
procedure is out of reach for the systems considered here, but we
expect the compromise to be very accurate, since PBE yields good
screening properties in the RPA, and HSE excellent approximations
for the true QP wavefunctions\cite{bulkGaN:Kr3}.

\begin{table}
 \caption{Lattice constant $a$, bulk-modulus $B_{0}$, energy gap at $\Gamma$, $L$, $X$,
dielectric constant $\varepsilon_{\infty}$, valence band-width $W$,
and  the energy position of Ga $d$ states determined using  PBE, HSE
and $GW_{0}$. Spin-orbit splitting is not included
($\Delta_{0}\approx~0.02$~eV). $\varepsilon_{\infty}$ was calculated
including local field effects and using the RPA (values in
parenthesis). } \label{tab:tab1}
\begin{ruledtabular}
\begin{tabular}{l|cccccc}
                & PBE & HSE & $GW_{0}$ & Exp \\
\hline
$a$(\AA)               & 4.546  &4.494 && 4.506$^{a}$\\
$B_{0}$(Gpa)         & 200    & 218  &&210 \\
E$_{\Gamma}$(eV)     &1.58    &3.06  &3.20& 3.26$^{b}$\\
E$_{L}$(eV)          &4.49    &6.18  &6.32&  \\
E$_{X}$ (eV)         &3.35    &4.47  &4.46&  \\
$\varepsilon_{\infty}$  &5.86 (5.55)    & 5.1 (4.6) && $\approx$ 5.3 \\ 
$W$(eV)                 & 7.1   & 7.9  &7.4     & 6.7$^{d}$ \\
$E_{d}$ &$-$13.3 &$-$15.3                 &$-$15.7& $-$17.0$^{e}$ \\
\end{tabular}
\end{ruledtabular}
$^{a}$ Ref.\cite{bulkGaN:ZincblendeGaN};$^{b}$
 Ref.\cite{bulkGaN:database1,bulkGaN:database2}; 
  $^{d}$ Ref.\cite{bulkGaN:width}; $^{e}$ Ref.\cite{bulkGaN:d}.
\end{table}
In Tab.\ \ref{tab:tab1} we report the  relevant equilibrium bulk
parameters of GaN in the zinc-blende
phase. The  HSE lattice constant is very close to the experimental value at
T=0$^\circ$ K. The E$_{\Gamma}^{\rm HSE}$ band-gap is 3.06 eV,
almost twice as large as the PBE one (1.58 eV), again in very good
agreement with experiment. Furthermore all HSE one-electron gaps are
very close to the $GW_{0}$ QP energies. A slight discrepancy arises
for the HSE  valence band-width, for which PBE gives the best
estimate (but experimental errors might be large). The position of
the Ga $d$ levels are  described well using HSE, only surpassed by
$GW_{0}$. Finally, the HSE  
dielectric constant agree well with experiment. Overall, the
agreement between experiment, HSE and $GW_{0}$ is very good.

\begin{figure}
\centering
\includegraphics[width=0.45\textwidth,angle=0,clip=true]{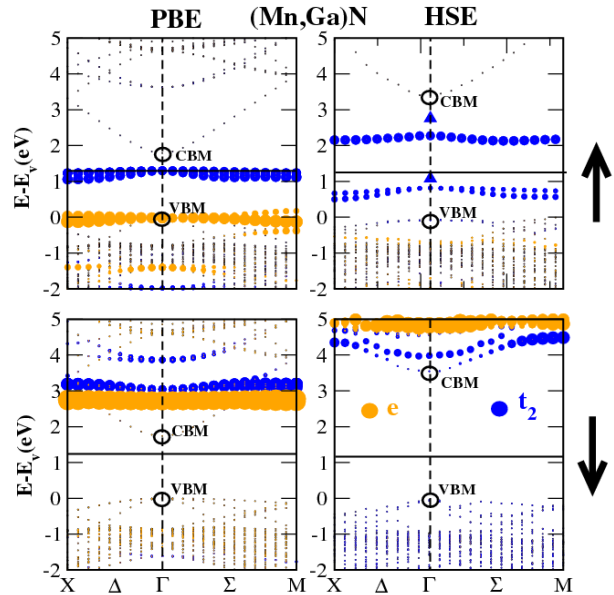}
\caption{(Color online) Spin-polarized projected band structure of
the Mn impurity in a cubic  64-atom cell using PBE (left) and HSE
(right). Majority (minority) bands are plotted in the upper (lower)
part. The horizontal line corresponds to the Fermi level (energy
zero is equivalent to  valence band maximum). The blue (dark gray) and orange (light gray)
circles indicate the strength of the $t_{2}$ (\emph{i.e.}
$d_{xy}$+$d_{xz}$+$d_{yz}$) and $e$ (\emph{i.e.}
$d_{x^{2}-y^{2}}$+$d_{3z^{2}-r^{2}}$) character, respectively. The
VBM and CBM of the host crystal are shown by small empty circles.
The majority doublet and singlet QP-shifts  are shown in the right
part by blue triangles.}\label{fig:bande64}
\end{figure}
Let us now consider the Mn substitution at the Ga host site.
 Fig. \ref{fig:bande64}
shows the band structure for the 64-atom cell plotted in the cubic
Brillouin zone along the symmetry lines  $\Delta$ and $\Sigma$.
In the PBE band-structure, the majority  t$_{2}$
states form an essentially dispersionless impurity band pinned at
the Fermi level. They are three-fold degenerate at $\Gamma$ with the
partial occupancy of each state equal to $2/3$. The three bands have
predominantly  $t_{2}$ character at the zone center ($\sim$0.39),
but they also show  some anti-bonding contributions from the N-$p$
states at the nearest N neighbors. The non-bonding $e$ states can be
found  at the valence band maximum
and they are strongly localized in the Mn sphere, with a total
$e$ character of $\sim$0.55. Below
the $e$ states, the GaN valence band states are found. They are
slightly hybridized with Mn states. For the minority component, the
$t_{2}$ and $e$ states are shifted above the host CBM due to
exchange splitting. The $e$ states form a flat band even more
localized than their majority counterpart 
($e$ character $\sim$0.80). 
The Fermi level  is located in the gap in the minority
component, while it cuts the $t_{2}$ bands in the majority states.
Therefore, in the PBE description, Mn$_{x}$Ga$_{1-x}$N  is a
half-metal  for a Mn concentration of $x\sim$3.1 \%.
 
The HSE results  differ significantly from the PBE results. First,
the band gap opening of the host is recognized. However, most
relevant is that the majority $t_{2}$ states are \emph{split} into
an unoccupied  singlet and an occupied doublet, which remains above
the host VBM. The energy separation of the singlet and doublet  is
1.46 eV at $\Gamma$, and the system is clearly insulating. 
The singlet state is strongly localized in the Mn sphere
($t_2$$\sim$0.40), whereas the doublet, which is close in energy
to the valence band, is less well localized, with a  $d$ character
of $\sim$0.20. The $e$ states do not form a flat band anymore, since
they are pushed down in energy hybridizing more effectively with the
host valence band. In the minority component, the $t_{2}$ band is
not as flat as in the PBE case (see Fig.\ \ref{fig:bande64}) but
hybridizes more strongly with the first host conduction bands,
especially at the zone boundaries, whereas the $e$ states remain
mostly unchanged, apart from a rigid upwards shift.

Remarkably, we observe that the quasiparticle $GW_{0}$ band structure is
essentially identical to the HSE one-electron energies. In the
$GW_0$ calculations, the doublet and the singlet shift upwards away
from the VBM by 0.25 eV and 0.43 eV compared to the HSE case, but
the energy separation remains almost unchanged. This suggests that
HSE is a legitimate shortcut for sophisticated many electron
calculations, an observation that already  transpires from the very
good HSE one-electron band-gaps. The $GW_{0}$ calculations also clearly
confirm that the metallic state observed in DFT-PBE is an artifact
of the involved approximations. In fact, a  band gap can be
predicted using DFT-PBE as well, if (and only if) the band gap is
calculated as the energy difference between the ionization potential
and electron affinity calculated by removing and adding {\em one}
electron in {\em one} supercell ($\sim$ 1 eV). However, {\em if a
single electron or hole were placed in a huge supercell with many Mn
substitutional sites}, DFT-PBE  predicts the electron affinity and
ionization potential to be equal (metal), 
whereas the hybrid functional predicts the band gap even in
the limit of low electron or hole concentrations. The latter result
is correct, whereas the PBE result is not in agreement with
experiment.
 
Another visible consequence of the splitting of the $t_{2}$ states
is the Jahn-Teller distortion around the defect. The Mn $t_{2}$
states do not  transform as an irreducible representation of the
T$_{d}$ group, since they are not 3-fold degenerate.
This effect is usually termed ''spatial symmetry
breaking"\cite{Broken1}. The splitting is even obtained at the
\emph{ideal} ionic structure, and for the ideal structure, three
equivalent electronic solutions with identical energies are found.
Each of the electronic solutions corresponds to a different charge
 ordering, and once the lattice is allowed to relax the
nuclear framework distorts accordingly, \emph{i.e.} the Jahn-Teller
effect is at play.

\begin{figure}
{\centering \includegraphics[width=0.3\textwidth,angle=0,clip=true]{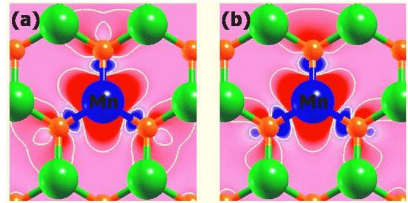}}
\caption{(Color online) PBE (a) and HSE (b)  spin-density
contours between -0.03 and 0.02 $\mu_{B}/$\AA$^{3}$
plotted on the (111) plane. Blue (dark gray) large spheres are Mn atoms; green (light gray) large
 spheres are Ga atoms;  orange (light gray) small spheres are N atoms. Red (dark gray) and blue (light gray)
 regions correspond to  positive and negative spin-density. Note the lack of the C$_{3}$ symmetry in (b).}
\label{fig:spindens}
\end{figure}

Fig.\ \ref{fig:spindens} shows the spin-density around the Mn atom
in the (111) plane in the ideal structure, for both PBE (a) and HSE
(b) calculated for a 128-atom cell, sampling the Brillouin zone at
the $\Gamma$ point only. The PBE spin-density is clearly symmetric,
while in the HSE case the symmetry is spontaneously broken with the
$C_{3}$ symmetry around all \{111\} directions missing. We can
determine the subgroup associated with the Jahn-Teller distortion,
by considering the character tables of  all the possible subgroups
of $T_{d}$\cite{symmetry} and taking into account that i) the
$t_{2}$ manifold is split into a singlet and a doublet, and ii) that
the subgroup must not contain the C$_{3}$ symmetry operation. The
only subgroup compatible with i) and ii) is D$_{2d}$.
By allowing the ions to relax without symmetry constraints,
 it is found that the tetrahedral environment around the Mn atom
 becomes distorted: two of the Mn-N bonds relax to 1.98\ \AA, while
 the other two bonds relax to 1.97\ \AA\ (the ideal  GaN distance is 1.94\ \AA). The relaxed
 structure is within numerical uncertainty  indeed consistent with
 the D$_{2d}$ symmetry, in agreement with
 experiments\cite{ExpJT1}.
 The energy gain due to
 relaxations is 184 meV/Mn. Semi-local functionals (PBE) are not able to capture the
 Jahn-Teller distortion \emph{unless} one introduces \emph{ad-hoc}
 symmetry breaking displacements, but even then the energy gain
 due to the symmetry breaking is certainly significantly underestimated\cite{JT-DFT}.

\begin{figure}
\centering
\includegraphics[width=0.45\textwidth,angle=0,clip=true]{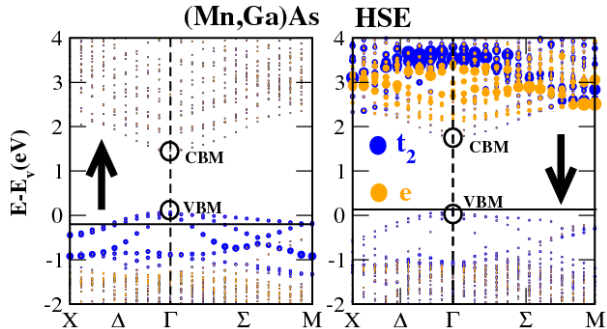}
\caption{(Color online) HSE spin-polarized projected band structure
of the Mn impurity in a cubic  GaAs 64-atom cell. See caption Fig.\ref{fig:bande64} for details.}\label{fig:bandeproj64}
\end{figure}

 As a final confirmation of the HSE results, we have calculated  the $\epsilon(0/-)$
 thermodynamic transition level for Mn$_{\rm Ga}$ in GaN by adding an electron in the
 HSE calculations and relaxing the geometry. We obtained a
 value of  1.9 eV comparing notably better with experiment (1.8$\pm$0.2~eV
 \cite{explevel})
 than PBE calculations (1.6~eV). In order to test the predictive capability of the procedure, 
we have repeated
 the calculations for Mn$_{\rm Ga}$ in GaAs, for which Mn is
 experimentally found
 in a $d^{5}$ configuration\cite{JT-DFT}. In this case,
 the splitting of the $t_{2}$ states
 drops to $\sim$ 0.2 eV, and no Jahn-Teller effect  is theoretically found
 using HSE or experimentally expected\cite{JT-DFT}. 
 Furthermore, the HSE band structure for Mn in
 GaAs (Fig.\ \ref{fig:bandeproj64}) shows that HSE
 predicts a metallic behaviour for the majority electrons. Hence 
 the hole introduced by Mn shows a transition from a \emph{localized} to
 a \emph{delocalized} character from GaN to GaAs, in full accordance
 with all experimentally available data.

A further confirmation of the accuracy of the
 present approach is that it is capable to predict the position of
 the Mn $d$ states, which  are  found at an average
 binding energy of 4.2 eV
 in photoemission experiments for GaAs\cite{dstates1},
 in very good agreement with the HSE results, while
 in SLDA or SGGA, the  spectral weight is shifted towards the Fermi level.

In summary, we have shown that the HSE functional predicts the 
localization of one electron hole at the  Mn impurity in GaN giving
rise to  a Jahn-Teller distortion and an insulating ground state via
splitting of the $t_{2}$ Mn manifold. The existence of a band gap
was confirmed by $GW_{0}$ quasiparticle calculations and is in agreement
with experiment. The formalism is able to predict the relevant
electronic features without imposing ad-hoc corrections on the $d$
states and includes ionic relaxations. Furthermore, a change of
the electronic hole state from a localized character in GaN to 
an itinerant band in GaAs is predicted. 
This observation  agrees with the fact that the localization is  strongly
environment dependent\cite{Nicola}. Previous interpretations of optical experiments\cite{opticrev} based on
standard DFT electronic
structure calculations
should be revised according to the picture emerging in this study.
Likewise, magnetic interactions should be re-investigated using functionals
that predict a band gap in the quasiparticle spectrum. Extension to IV-group 
DMS are currently in progress.\cite{unpublished}

\acknowledgments This work was supported by the Austrian {\em Fonds
zur F\"orderung der wissenschaftlichen Forschung}. A.S. thanks K.
Hummer for technical assistance in the calculation of dielectric
properties.

\end{document}